\documentclass[ALICE,manyauthors]{cernphprep}

\usepackage[comma,square,numbers,sort&compress]{natbib}
\usepackage{hyperref}
\usepackage{lineno}

\begin{document}
%
%
\begin{titlepage}
\PHyear{2015}
\PHnumber{196}                 
\PHdate{23 July}              
%
%
\title{Study of cosmic ray events with high muon multiplicity using the ALICE detector at the CERN Large Hadron Collider}
\ShortTitle{Atmospheric muons in ALICE}   
%
\Collaboration{ALICE Collaboration%
         \thanks{See Appendix~\ref{app:collab} for the list of collaboration
                      members}}
\ShortAuthor{ALICE Collaboration}      
%

\begin{abstract}
ALICE is one of four large experiments at the CERN Large Hadron Collider near Geneva, 
specially designed to study particle production in ultra-relativistic heavy-ion collisions.  
Located 52 meters underground with 28 meters of overburden rock, it has also been used to detect 
muons produced by cosmic ray interactions in the upper atmosphere. In this paper, we present the 
multiplicity distribution of these atmospheric muons and its comparison with Monte Carlo simulations.
This analysis exploits the large size and excellent tracking capability 
of the ALICE Time Projection Chamber.
A special emphasis is given to the study of high multiplicity events containing more than 100 reconstructed
muons and corresponding to a muon areal density $\rho_{\mu} > 5.9~$m$^{-2}$. 
Similar events have been studied in previous underground experiments such as ALEPH 
and DELPHI at LEP.  While these experiments were able to reproduce the measured muon multiplicity distribution with Monte Carlo simulations at low and intermediate multiplicities, their simulations failed to describe the 
frequency of the highest multiplicity events.  
In this work we show that the high multiplicity events observed in ALICE stem from primary cosmic rays with energies above $10^{16}$~eV and that the frequency of these events can be successfully described by assuming a heavy mass composition of primary cosmic rays in this energy range.
The development of the resulting air showers was simulated using the latest
version of QGSJET to model hadronic interactions. 
This observation places significant constraints on alternative, more exotic, production mechanisms for these events.
\end{abstract}

\end{titlepage}

\newpage
\setcounter{page}{2}
\section{Introduction}

ALICE (A Large Ion Collider Experiment) \cite{alice_exp} designed to study Quark-Gluon Plasma (QGP) formation in ultra-relativistic heavy-ion collisions at the CERN Large Hadron Collider (LHC), has also been used to perform studies that are of 
relevance to astro-particle physics.
The use of high-energy physics detectors 
for cosmic ray physics was pioneered by ALEPH \cite{aleph}, DELPHI \cite{delphi}
and L3 \cite{l3c} 
during the Large Electron-Positron (LEP) collider era at CERN.  An extension of these earlier studies is 
now possible at the LHC, where experiments can operate under stable conditions for many years.  ALICE undertook a programme of cosmic ray data taking between 2010 and 2013 during pauses in collider operations when there was no beam circulating in the LHC.  

Cosmic ray muons are created in Extensive Air Showers (EAS) following the interaction of cosmic ray
primaries (protons and heavier nuclei) with nuclei in the upper atmosphere. Primary cosmic 
rays span a broad energy range, starting at approximately $10^{9}$ eV and extending to more than $10^{20}$ eV. 
In this study, we find that events containing more than four reconstructed muons in the ALICE Time Projection Chamber (TPC), which we refer to as 
{\it multi-muon events}, stem from primaries with energy $E>10^{14}$ eV.  The detection of EAS originating from interactions above this energy, in particular around the energy of the knee in the primary spectrum ($ E_{k} \sim 3 \times 10^{15}$ eV), 
has been performed by several large-area arrays at ground level  
(e.g. \cite{eastop, casa, kascade}), 
while deep underground detectors 
(e.g. \cite{nusex, macro, emma}) 
have studied the high energy muonic component of EAS. 
The main aims of these experiments were to explore the mass composition and energy spectrum of primary 
cosmic rays. 

The muon multiplicity distribution (MMD) was measured at LEP with the ALEPH 
detector~\cite{aleph_mmd}.  This study concluded that the bulk of the data can be successfully described
using standard hadronic production mechanisms, but that the highest multiplicity events, containing
around 75-150 muons, occur with a 
frequency which is almost an order of magnitude above expectation, even when assuming that the primary 
cosmic rays are purely composed of iron nuclei.  
A similar study was carried out with the DELPHI detector, which also found that Monte Carlo simulations 
were unable to account for the abundance of high muon multiplicity events~\cite{delphi_mmd}.  
Several proposals have been put forward in the scientific literature to explain this discrepancy.  
Some authors suggest that hypothetical strangelets form a small percentage of very energetic cosmic 
rays~\cite{strangelets}, while others have tried to explain the excess of high muon multiplicity events
by the creation of the QGP in interactions involving high mass primary cosmic rays (iron nuclei) 
with nuclei in the atmosphere~\cite{ridky}. 
   
In this paper, we exploit the large size and excellent tracking capability of the ALICE TPC~\cite{TPC} to study the muonic component of EAS.  We describe the analysis of the muon
multiplicity distribution with particular emphasis on high muon multiplicity events containing more than 
100 muons in a single event and corresponding to an areal density $\rho_{\mu} > 5.9$~m$^{-2}$.  
We employ a description of the shower based upon the latest version of QGSJET~\cite{qgsjet1,qgsjet2}, a hadronic interaction model commonly used in EAS simulations.
  
Details of the environment of ALICE and the detectors used for this analysis are described in the following 
section, while the selection of the data and the algorithm adopted to reconstruct atmospheric muons are 
discussed in Section 3. The muon multiplicity distribution and the study of high muon multiplicity 
events are described in Section 4. The results are presented in Section 5 and, finally, in Section 6 we make 
some concluding remarks.   

\section{The ALICE experiment}

ALICE is located at Point 2 of the LHC accelerator complex, approximately 450 m above sea level in a cavern 52 m underground with 28 m of overburden rock.  The rock absorbs all of the electromagnetic  and hadronic components of the observed EAS,  so that only muons with an energy $E$, at the surface, larger than $16$~GeV reach the detectors \cite{alice_ppr2}.  The geometry of ALICE is typical of a collider 
experiment. A large solenoidal magnet forms a central barrel that houses several detectors, including a large, cylindrical TPC.  Outside the solenoid, and on one end, there is a single-arm, forward spectrometer, 
which was not used in this analysis.  A complete description of the apparatus is given in \cite{alice_exp}. 

\begin{figure}[!ht]
\centering 
\includegraphics[width=0.8 \textwidth]{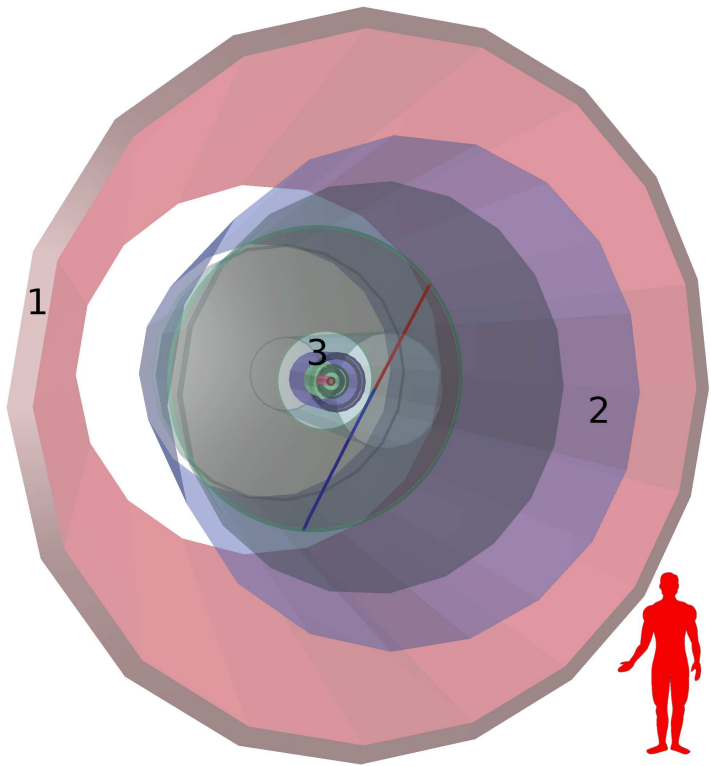}
\caption{A single atmospheric muon event. The thin outer cylinder is the Time Of Flight detector (1).  
The large inner cylinder is the Time Projection Chamber (2) and the smaller cylinder at the centre is 
the silicon Inner Tracking System (3).  Muons are reconstructed as two TPC tracks, one in the upper half of 
the detector  (\emph{up} track) and the other in the lower half (\emph{down} track), which are then joined 
to create a single muon track.}
\label{fig:muontracks}
\end{figure}

The ALICE TPC is the largest detector of its type ever built. It was used to reconstruct 
the trajectory of cosmic ray muons passing through the active volume of the detector, which comprises
a cylindrical gas volume divided into two halves by a central membrane.
The TPC has an inner radius of 80~cm, an outer radius of 280~cm and a total length of 500~cm along the LHC beam direction.   At each end of the cylindrical 
volume there are multi-wire proportional chambers with pad readout.  
For the purpose of detecting cosmic ray muons, the total area of the detector due to its horizontal cylindrical geometry is approximately 26~m$^2$.  
However, after placing a cut on the minimum length required to reconstruct a cosmic ray track the maximum effective area reduces to approximately 17~m$^2$.  
The apparent area of the detector also varies with the zenith angle of the incident muons.
Track selection is discussed in more detail in Section~\ref{sec:reco}. 
An example of a single atmospheric muon event is shown in Fig.~\ref{fig:muontracks}.

Three 
detector subsystems were used to provide dedicated triggers for this study: ACORDE 
(Alice COsmic Ray DEtector)~\cite{ACORDE}, SPD (Silicon Pixel Detector)~\cite{SPD} and TOF 
(Time Of Flight detector)~\cite{TOF}. 

ACORDE is an array of 60 scintillator modules located on the three upper faces of the octagonal yoke of the solenoid, 
covering $ 10\% $ of its surface area. A trigger was formed by the coincidence of signals in two different 
modules (a two-fold coincidence), although the trigger can also be configured to select events when a single 
module fires or when more than two modules fire. 

The SPD is part of the Inner Tracking System located inside the inner field cage of the TPC.  
It is composed of two layers of silicon pixel modules located at a distance of 39 mm and 76 mm from 
the LHC beam axis, respectively.  The layers have an active length of 28.3 cm, centred 
upon the nominal interaction point of the LHC beams. The SPD was incorporated into the trigger by 
requiring a coincidence between signals in the top and bottom halves of the outermost layer. 

The TOF is a cylindrical array of multi-gap resistive-plate chambers that completely surrounds the outer radius of
the TPC.  
The TOF trigger requires a signal in a pad corresponding to a cluster of readout channels covering 
an area of 500\ cm$^2$ in the upper part of the detector and another signal in a pad in the opposite lower part
forming a back-to-back coincidence with respect to the central axis of the detector.
The configuration of the pads involved in the trigger can be changed via software. 
In some periods of data taking, this flexibility has been exploited to require a signal in an upper pad and in the 
opposing pad plus the two adjacent pads forming a back-to-back $\pm$1 coincidence.

Cosmic ray data were acquired with a combination (logical OR) of at least two out of the three trigger
conditions (ACORDE, SPD and TOF) depending on the run period.
The trigger efficiency was studied with a detailed Monte Carlo simulation, which is discussed in 
Section \ref{simulation}.  
Most events were classified as either single muon events or multi-muon events, 
with a small percentage of ``interaction" events where very energetic muons have interacted with the iron 
yoke of the magnet producing a shower of particles that pass through the TPC. 

\section{Event reconstruction and data selection}
\label{sec:reco}

The TPC tracking algorithm~\cite{Performance} was designed to reconstruct tracks produced in  
the interaction region of the two LHC beams.  It finds tracks by working inwards from the outer radius of the 
detector where, during collider operation, the track density is lowest.  The present analysis used the same tracking algorithm 
but removed any requirement that tracks should pass through a central interaction point.  However, the
tracking algorithm has not been optimised for very inclined (quasi horizontal) tracks.  Therefore, to avoid 
reconstruction inaccuracies associated with the most inclined showers, we restricted the zenith angle 
of all events to the range $0^{\circ}<\theta<50^{\circ}$.  

As a consequence 
of reconstructing tracks from the outer radius of the TPC inwards, cosmic ray muons are typically 
reconstructed as two separate tracks in the upper and lower halves of the TPC as shown in 
Fig.~\ref{fig:muontracks}.  We refer to these tracks as \emph{up} and \emph{down} tracks.  Following this first 
pass of the reconstruction a new algorithm was applied to match each \emph{up} track with its 
corresponding \emph{down} track to reconstruct the full trajectory of the muons and to eliminate 
double counting. Starting with single muon events (producing two TPC tracks), where the matching of tracks is 
straightforward, the reconstruction has been tuned to handle events containing hundreds of muons.  
High multiplicity Monte Carlo events have been used to optimise the matching performance.

Each TPC track can be reconstructed with up to 159 individual space points.  In order to maximise the
detector acceptance for this analysis, tracks were required to have a minimum of 50~space points and, 
in events where the magnetic field was on, a momentum greater than 0.5 GeV/$c$ 
to eliminate all possible background from electrons and positrons. 
In multi-muon events, accepted tracks were required to be approximately parallel since
atmospheric muons coming from the same EAS arrive almost parallel at ground level. 
The parallelism cut involves forming the scalar product of the direction 
of the analysed track $\vec{t_a}$ with a reference track 
$\vec{t_r}$, requiring that $ \vec{t_a} \cdot \vec{t_r} = \cos (\Delta\Psi) > 0.990$ 
to accept the analysed track.  The reference track was 
chosen to give the largest number of tracks satisfying the parallelism cut. 
This requirement introduces an additional momentum cut due to the bending of muon tracks in the
magnetic field. The momentum cut is a function of the azimuth angle of the muon track and varies
between 1 and 2 GeV/$c$.  
Finally, each up track was matched to the nearest 
down track if the distance of closest approach between them at the horizontal mid plane 
of the TPC was $d_{xz}<3$ cm. This value was chosen to be large enough to maximise the matching 
efficiency in high multiplicity Monte Carlo events, while keeping combinatorial background to a minimum.

A muon reconstructed with two TPC tracks (up and down) is called a ``matched muon".   When a TPC 
track fulfils all the criteria to be a muon track: number of space points, momentum and parallelism, but does not 
have a corresponding track within $d_{xz}<3$\ cm in the opposite side of the TPC, this track is still accepted 
as a muon candidate but flagged as a ``single-track muon". Most single-track muons are found to cross 
the TPC near its ends where part of the muon trajectory falls outside the detector. 

\begin{figure}[!ht]
  \centering 
  \includegraphics[width=0.8 \textwidth]{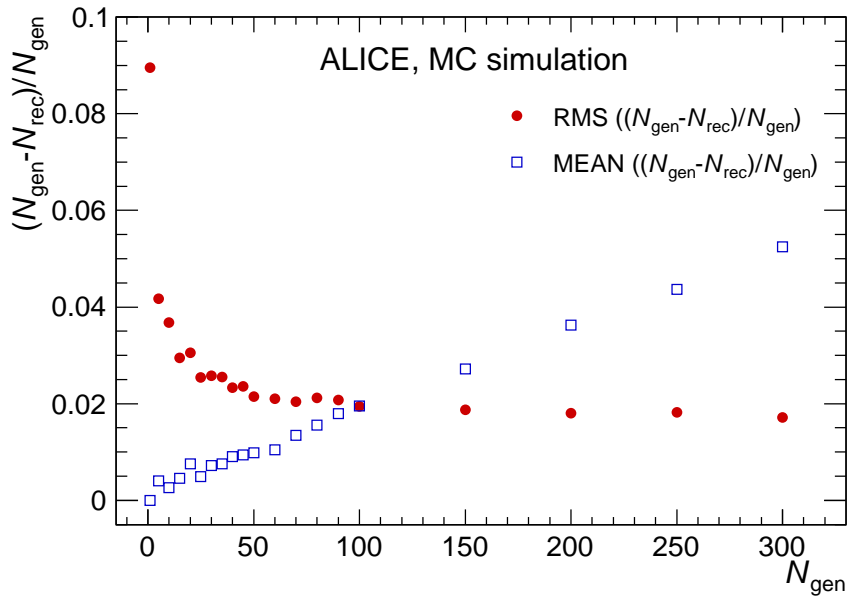}
  \vspace{2ex}
  \caption{Root-mean-square and mean values of the relative difference between the number of generated and reconstructed muons for events simulated with different muon multiplicities.}
  \label{fig:rms_deltanmu}
\end{figure}

To quantify the performance of the tracking and matching algorithms, we studied the multiplicity
dependence of the reconstruction efficiency using Monte Carlo simulated events.  We generated 
1000 events for 20 discrete values of the muon multiplicity, varying between 1 and 300, which were then reconstructed using the same algorithms applied to real events.  
In each event, muons were generated parallel to each other like in EAS and cross the whole TPC volume.
Fig.~\ref{fig:rms_deltanmu} shows the mean values (MEAN) and root-mean-square (RMS) of
the relative difference between the number of generated 
and reconstructed muons,
\begin{equation}
\left(\#~\text{generated muons} - \#~\text{reconstructed muons}\right) / \left(\#~\text{generated muons}\right),
\end{equation}
as a function of the number of generated muons.
The root-mean-square represents the resolution on the number of reconstructed muons and is typically less than $4\%$, while for the highest multiplicities it is around $2\%$. The mean value is less than $1\%$ up to $N_{\mu} \approx 50$, increasing to $5\%$ at high muon multiplicities ($N_{\mu} \approx 300$). 

\begin{figure}[!ht]
  \centering 
  \includegraphics[width=0.8 \textwidth]{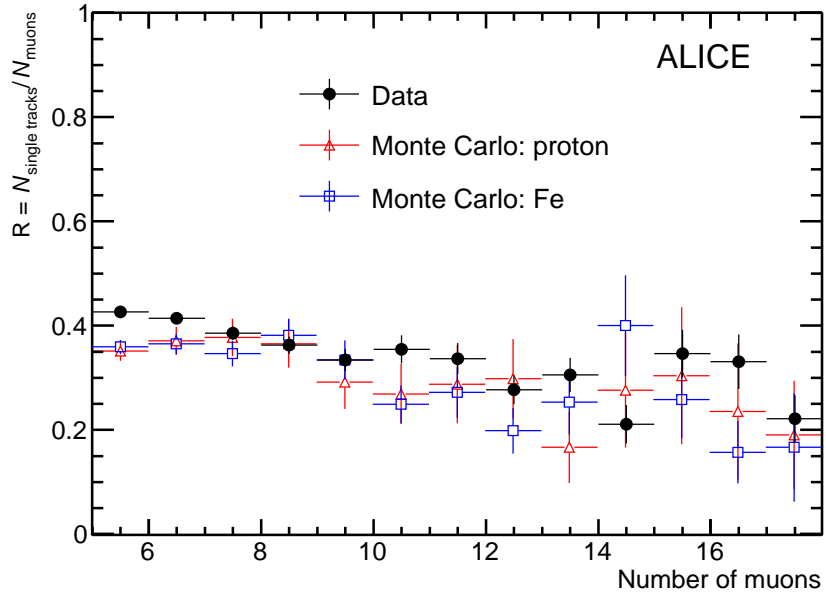}
  \vspace{2ex}
  \caption{The ratio of muons reconstructed as single tracks to the total number of reconstructed muons (both single and matched tracks) in the data and simulations with proton and iron primaries.}
  \label{fig:single_matched}
\end{figure}

To illustrate the similarity of the data and the Monte Carlo simulation, 
Fig.~\ref{fig:single_matched} shows the ratio of the number of muons reconstructed as single tracks
(either {\it up} or {\it down} tracks) to the total number of reconstructed muons (both single and matched tracks)
for different multiplicities. The ratio obtained from the data is compared with the ratios 
obtained from simulated samples of pure proton primary cosmic rays and pure iron primaries.
Over the range of intermediate muon multiplicities shown, the ratio varies between 0.2 and 0.4 with 
good agreement between data and simulations.  There is no significant difference between the simulated
proton and iron samples.      
 
Data were recorded
between 2010 and 2013 during pauses in collider operations when no beam was circulating in the LHC.  
The total accumulated run 
time amounted to 30.8 days,
resulting in approximately 22.6~million events with at least one
reconstructed muon (single-track or matched) in the TPC. 
Only multi-muon events are discussed further in this paper.  We define 
{\it multi-muon events} as those events with more than four reconstructed muons in the TPC 
($N_{\mu}>4$).  In total, we collected a sample of 7487 multi-muon events.
 
\section{Analysis of the data and simulation}
\label{simulation}

\begin{figure}[!ht]
\centering 
\includegraphics[width=0.7 \textwidth]{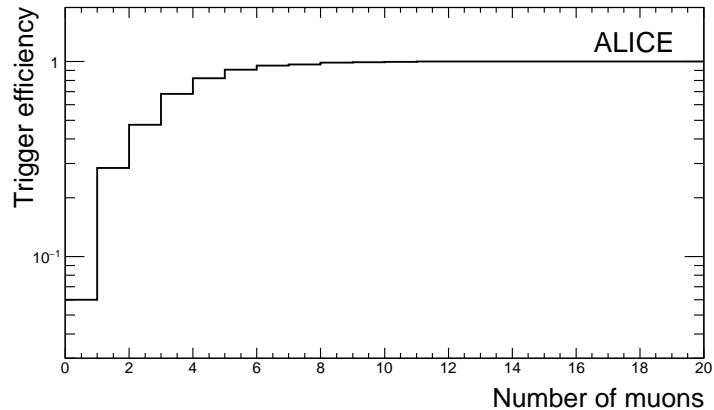}
\caption{TOF trigger efficiency as a function of muon multiplicity.}
\label{fig:tof_eff}
\end{figure}

To obtain the MMD we have corrected the measured distribution for the efficiency of the trigger. 
The correction was calculated from a Monte Carlo simulation that is described later in this section.  
Given the complementary coverage of the TOF barrel to the TPC, the TOF trigger was 
mainly responsible for selecting events in the low-to-intermediate range of muon multiplicities 
($7~\leq~N_{\mu}~\leq~70$).  The efficiency of the TOF trigger as a function of the muon multiplicity 
is shown in Fig.~\ref{fig:tof_eff}.  The efficiency is lower at low muon multiplicity due to the 
back-to-back coincidence requirement of the TOF trigger. The efficiency of the ACORDE trigger has
a similar, increasing trend with the muon multiplicity.  The multiplicities at which the two triggers 
reach full (100\%) efficiency are $N_{\mu} > 10$ (TOF) and $N_{\mu} > 15$ (ACORDE).  
Given the much smaller area of the SPD in comparison with the TPC, the efficiency of the SPD trigger
is significantly lower than both ACORDE and TOF. It makes only a minor contribution to the MMD in the low-to-intermediate range of muon multiplicities. 

\begin{figure}[!ht]
\centering 
\includegraphics[width=0.8 \textwidth]{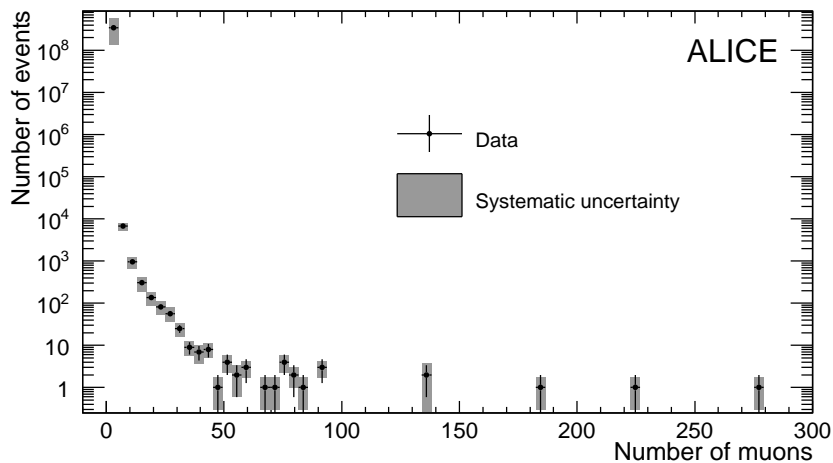}
\caption{Muon multiplicity distribution of the whole sample of data (2010-2013) corresponding to 30.8 days of data taking.}
\label{fig:mmd_data_31d}
\end{figure}

The MMD obtained from the whole data sample and corrected for trigger efficiency is shown in 
Fig.~\ref{fig:mmd_data_31d}. Values for the systematic uncertainty in the number of events as a function 
of multiplicity have been estimated by varying the parameters of the track reconstruction and matching 
algorithms. We find a smooth distribution up to a muon multiplicity of around 70 and then 5 events with 
a muon multiplicity greater than 100.
We define the events with $N_\mu>100$ high muon multiplicity (HMM) events. Given the nature and topology of 
high multiplicity events, all trigger conditions contributed to this sample with close to 100\% efficiency.
The aim of the following analysis is to model the MMD at low-to-intermediate multiplicities and to explore the 
origin of the HMM events.

The difficulty in describing EAS, and consequently the number of muons reaching ground level, mainly arises 
from uncertainties in the properties of multi-particle production in hadron-air interactions. 
These interactions are often described phenomenologically within Monte Carlo event generators.
Model parameters, such as total and inelastic hadron-proton cross sections, inelastic scattering slopes 
and diffractive structure functions, are constrained by measurements obtained from accelerator experiments. 

In this analysis we have adopted the CORSIKA~\cite{corsika} event generator incorporating 
QGSJET~\cite{qgsjet1} for the hadronic interaction model to simulate the generation and development of EAS. 
CORSIKA version 6990 incorporating QGSJET II-03 has been used to study the MMD distribution and
HMM events; CORSIKA version 7350 incorporating QGSJET II-04 has been used to check and 
confirm the results for HMM events.  
The significant differences between the two versions of QGSJET are the inclusion of Pomeron loops in 
the formalism of QGSJET II-04 and a retuning of the model parameters using early LHC data for the 
first time~\cite{qgsjet_2011}.  Most relevant to the present study is that pion exchange is assumed to 
dominate forward neutral hadron production in the QGSJET II-04, which has been shown to enhance 
the production of  $\rho^0$ mesons resulting in an enhancement of the muon content of EAS by about 20\%~\cite{epjweb_2013}.

In previous studies of cosmic ray muon events at LEP, QGSJET 01 was used to model hadronic
interactions.  Apart from the way in which nonlinear effects are modelled,
another significant difference between this earlier version of the model and QGSJET II-03/04
is the deeper shower maximum, $X_{max}$, used in the later versions.  This results in a steeper lateral muon 
distribution and an associated increase of the muon density close to the core of the shower,
which can also have an impact on the observed rate of HMM events.

When generating cosmic ray events, the core of each shower was scattered randomly at ground level 
over an area covering $205 \times 205$ m$^{2}$ centred upon the nominal LHC beam crossing point.  
This area was chosen to minimise the number of events to be generated without creating any
bias on the final results.
We found that, when the core was located outside this area, only a very small number of events 
gave rise to muons crossing the TPC and these events were always of low multiplicity 
($N_{\mu} < 4$).  Therefore, neglecting these events does not affect the results reported in this paper. 

To have a fast and flexible method of estimating several important parameters and observables
involved in the analysis, we started with a simplified Monte Carlo simulation. This simulation 
did not explicitly model interactions in the rock above the experiment. Instead, the trajectories of the muons 
arriving at the surface were simply extrapolated as straight lines to the depth of ALICE while imposing
an energy cut $E_{\mu} > 16~\text{GeV} / \cos(\theta)$, where $\theta$ is the zenith angle of the muon.
All muons passing this cut and crossing an area of 17~m$^2$, corresponding to the horizontal cross-sectional 
area of the TPC, were considered to be detected. 

To understand the complete sample of the recorded data, including the origin of low muon multiplicity 
events, we generated events initiated by the interaction of proton and iron ($^{56}$Fe) 
primaries with energies $E > 10^{12}$ eV. 
This revealed that most single muon events stem from primaries in the energy range 
$10^{12} < E < 10^{13}$ eV, while primaries in the energy range $10^{13} < E < 10^{14}$ eV 
produce muon multiplicities typically in the range from 1 to 4, independent of the mass of
the primary cosmic rays.
Primaries with energies below $10^{14}$~eV therefore produce a negligible 
contribution to multi-muon events ($N_{\mu}>4$) that are of interest in this study. 
Consequently, only energies $E > 10^{14}$~eV were considered in the full simulation.
 
The first step in the analysis was to attempt to reproduce the measured MMD 
in the low-intermediate range of multiplicity ($7 \leq N_{\mu} \leq 70$). 
Samples of proton and iron primary cosmic rays were generated in the energy range 
$10^{14} < E < 10^{18}$ eV and with zenith angles in the interval $0^{\circ}<\theta<50^{\circ}$. 
The composition of cosmic rays in this energy range is a mixture of many species of nuclei 
in a ratio that is not well-known and which varies with energy. To simplify the analysis and interpretation 
of the data we have modelled the primary cosmic ray flux using a pure proton sample, 
representing a composition dominated by light nuclei, and a pure iron sample, representing
a composition dominated by heavy nuclei. In relation to the MMD, the proton sample provides a lower limit
on the number of events for a given multiplicity, while the iron sample provides an upper limit.
A typical power law energy spectrum,  $E^{-\gamma}$, has been adopted 
with a spectral index $\gamma = 2.7 \pm 0.03$ for energies below the knee ($E{_k}=3 \times 10^{15}$ eV) and    
$\gamma_k = 3.0 \pm 0.03$ for energies above the knee. The total (all particle) flux of cosmic rays has been calculated by summing the individual fluxes of the main chemical elements at 1 TeV \cite{horandel_flux_cr} 
where measurements are most precise. 
The flux was estimated to be $F(1\ $TeV$) = 0.225 \pm 0.005$ (m$^2$ s sr TeV)$^{-1}$.

All events generated with energies $E>10^{14}$ eV 
were subsequently considered for a complete analysis using a detailed
simulation taking into account all possible interactions in matter surrounding the experiment.
In each event, all muons were extrapolated to the horizontal mid-plane of the experiment
and flagged if they hit an enlarged area of 36~m$^2$ centred upon the TPC with no
restriction on the energy of the muons.  
All flagged muons were recorded along with their position and momentum at ground level 
and used as input to the ALICE simulation framework. 
In this framework, the ALICE experimental hall and the environment above and around 
the apparatus as well as all the detectors are accurately described.  
Flagged muons were propagated through this environment with GEANT3~\cite{geant3}.  
Any muon that crossed the detector apparatus was treated by a detector response simulation 
that produced pseudo-raw data, which was then processed with the same reconstruction code
that was applied to real data, including the TPC tracking algorithm and the track matching algorithm 
developed for this analysis. 

\subsection{The muon multiplicity distribution}
\label{sec:mmd}

\begin{figure}[!ht]
\centering 
\includegraphics[width=0.8 \textwidth]{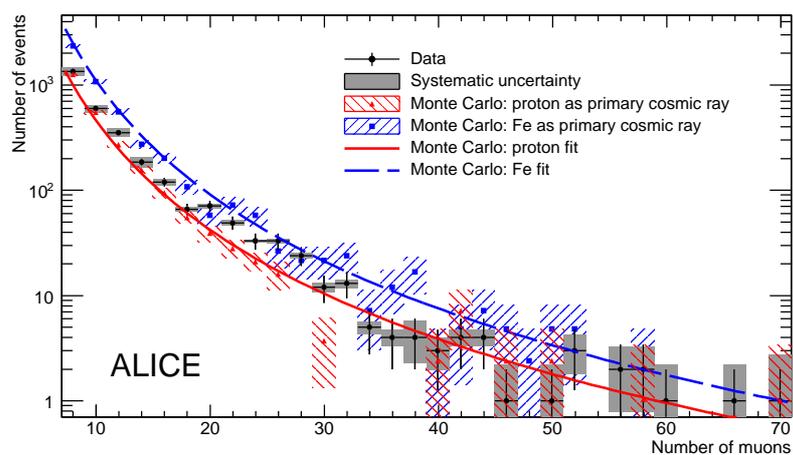}
\caption{The measured muon multiplicity distribution compared with the
values and fits obtained from CORSIKA simulations with proton and 
iron primary cosmic rays for 30.8 days of data taking. The errors are shown separately
(statistical and systematic) for data, while for Monte Carlo they are the quadrature sum of the
statistical and systematic uncertainties. }
\label{fig:mmd_data_p_fe_31d}
\end{figure}

We generated simulated events equivalent to 30.8 days live time to permit direct comparison with the data without the need to apply an arbitrary normalisation factor.
A comparison of the trigger corrected, measured MMD with the simulations is shown in 
Fig.~\ref{fig:mmd_data_p_fe_31d}. For ease of comparison, the points 
obtained with the simulations were fitted with a power-law function to 
obtain the curves for proton and iron.

At lower multiplicities, corresponding to lower primary energies, we find that the data 
approach the proton curve, which represents a light ion composition of the primary cosmic ray flux, 
while higher multiplicity data lie closer to the iron curve, representing a heavier composition.  
The limited statistics in the range $N_{\mu} > 30$ does not allow for a precise, 
quantitative study of the composition but suggests that the average mass of the primary 
cosmic ray flux increases with increasing energy, a finding consistent with several previous 
experiments~\cite{eastop_2004,casamia_1999,macroeastop_2004,kascade_2005}.

The errors in Fig.~\ref{fig:mmd_data_p_fe_31d} are shown separately (statistical and systematic) for data, while for Monte Carlo they are the quadrature sum of the
statistical and systematic uncertainties.
The systematic errors in the simulations take into account
uncertainties in the flux of cosmic rays at 1 TeV, the slope of the energy spectrum below and above 
the knee, the description of the rock above the experiment and the uncertainty in the 
the number of days of data taking (detector live time).   
The largest contribution to the systematic error is due to the uncertainty in the spectral index 
below the knee ($\gamma = 2.7 \pm 0.03$), which results in an uncertainty of approximately $15\%$
in the MMD.  
The error in the description of the rock above the experiment corresponds to an uncertainty 
in the energy threshold of the muons reaching the detector, which results in a systematic error 
of approximately $4\%$.
Each of the other uncertainties gives a contribution of around $2\%$ to the systematic error. 
For muon multiplicities $N_{\mu} > 30$, statistical uncertainties are dominant.

Following success in describing the magnitude and shape of the MMD over this 
intermediate range of multiplicities ($7 \leq N_{\mu} \leq 70$) we have used the same simulation 
framework to study the frequency of HMM events.  Since these are particularly rare events, a very 
high statistics sample of simulated HMM events was required to permit a meaningful quantitative comparison. 

\subsection{High muon multiplicity events}
\label{sec:hmm}

\begin{figure}[!ht]
\centering 
\includegraphics[width=0.8 \textwidth]{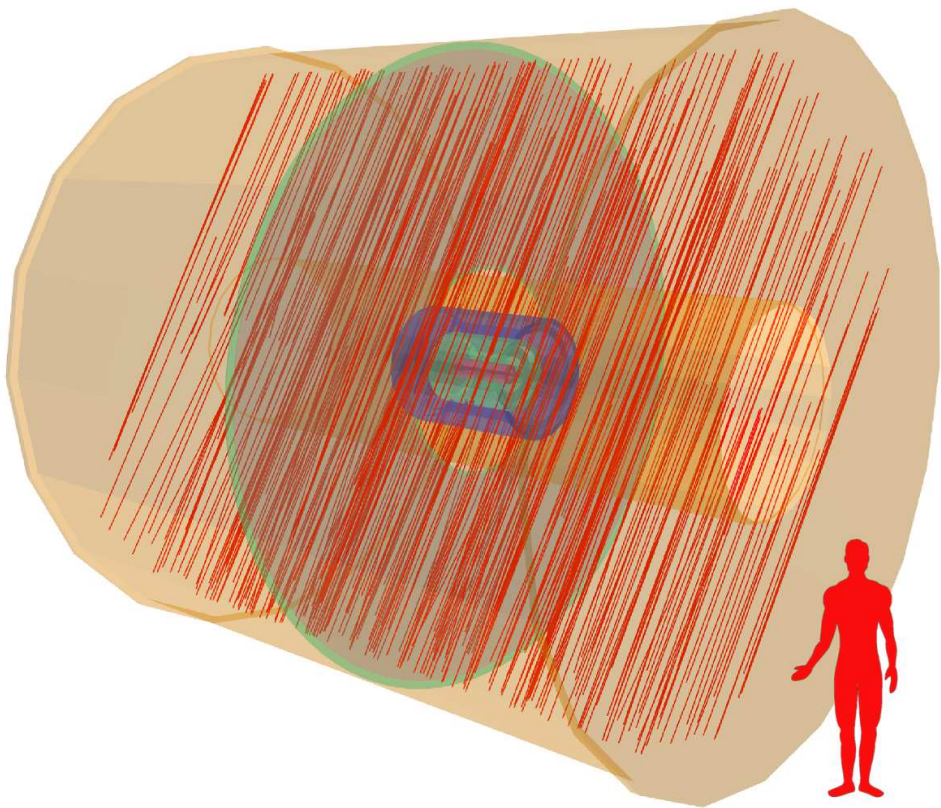}
\caption{Event display of a multi-muon event with 276 
reconstructed muons crossing the TPC.}
\label{fig:evdisplay_276}
\end{figure}

\begin{figure}[!ht]
\centering 
\includegraphics[width=0.8 \textwidth]{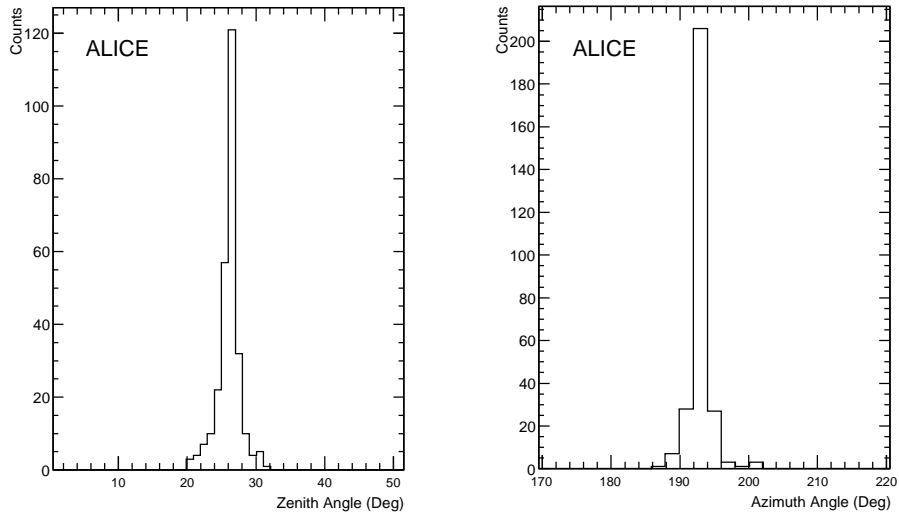}
\caption{Zenithal and azimuthal distribution of the multi-muon event 
with 276 reconstructed muons.}
\label{fig:evangdist_276}
\end{figure}

\begin{figure}[!ht]
\centering 
\includegraphics[width=0.8 \textwidth]{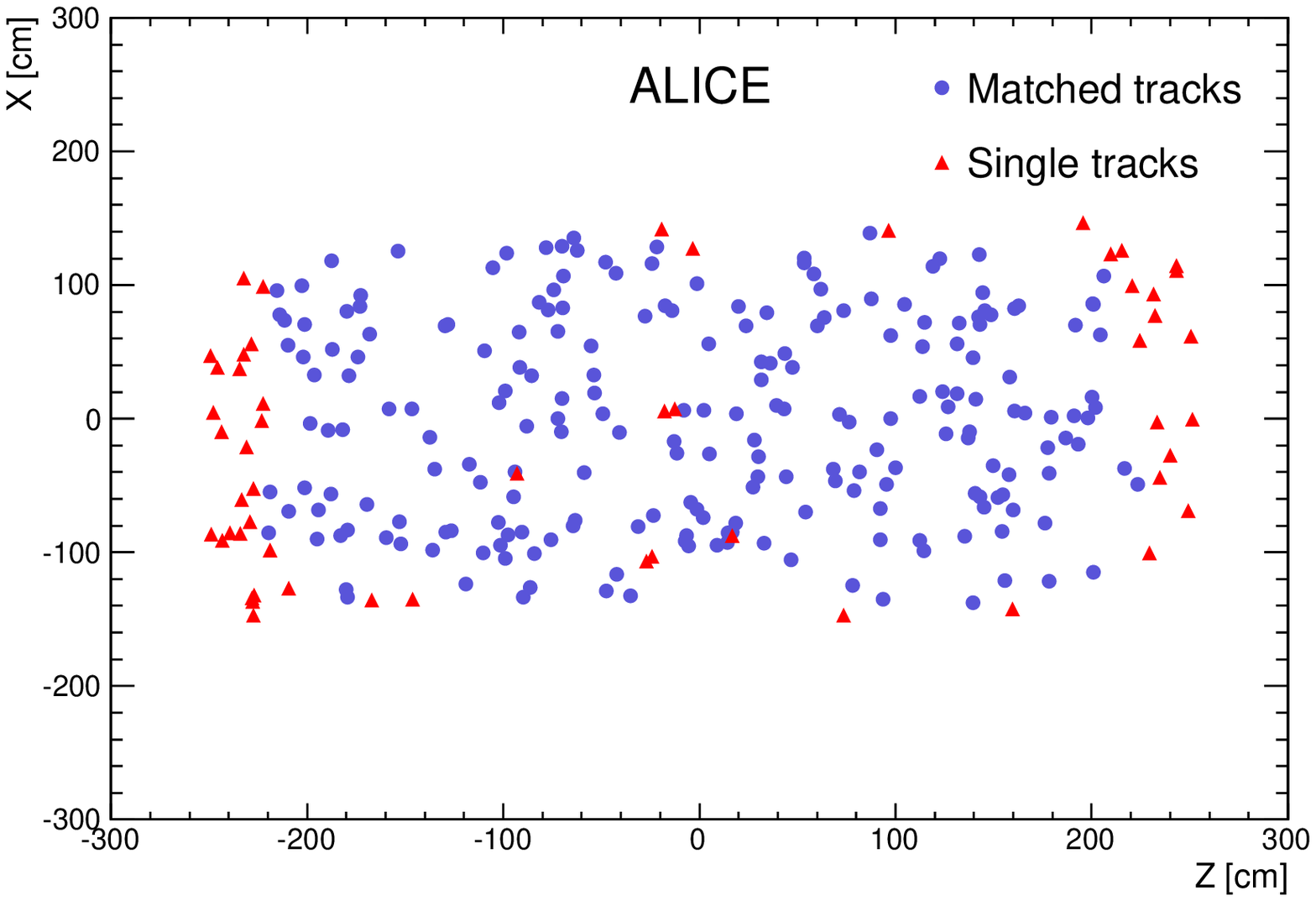}
\caption{Spatial distribution of the 276 recostructed muons indicating
matched and single-track muons.}
\label{fig:evspatdist_276}
\end{figure}

Taking the dataset as a whole, corresponding to 30.8 days and a mixture of running conditions, we find 5 HMM events with muon multiplicities $N_{\mu}>100$ 
(as can be seen in Fig.~\ref{fig:mmd_data_31d}) 
giving a rate of $1.9 \times 10^{-6}$~Hz.  Each of these events 
were examined closely to exclude the
possibility of ``interaction'' events.  The highest multiplicity event
reconstructed in the TPC was found to contain 276 muons, which corresponds to 
a muon areal density of 18.1~m$^{-2}$. 
For illustration, a display of this event is shown in Fig.~\ref{fig:evdisplay_276}. 
The zenithal and azimuthal angular distributions of the muons from the same HMM event are shown in 
Fig.~\ref{fig:evangdist_276}, while the spatial distribution of matched and single-track muons at the TPC
mid plane is shown in Fig.~\ref{fig:evspatdist_276}. We note that the majority of single-track muons are 
reconstructed near the ends of the TPC  where muons may enter or leave the active volume
without producing a track either the upper or lower halves of the detector.

One of the aims with this study is to compare the rate of HMM events obtained from simulations to the measured rate.
To limit the effect of fluctuations in the number of simulated HMM events, we have simulated a live time 
equivalent to one year with CORSIKA 6990 using QGSJET II-03 for the hadronic interaction model.  
The simplified Monte Carlo used as a first step of the analysis demonstrated that only primaries with energy 
$E>10^{16}$~eV contribute to these events. Therefore, only events in the range of primary energy 
$10^{16}<E<10^{18}$ eV have been generated to achieve an equivalent of 365 days exposure for both proton 
and iron primaries.

The estimated maximum fiducial area of the TPC due to 
its horizontal cylindrical geometry and cut on the minimum number 
of TPC space points is $17 \pm 0.5$~m$^2$.  The estimated error in the number of reconstructed muons, $N_\mu$,
counting both matched and single-track muons, is around $5\%$ for $N_\mu>100$.  HMM events are therefore events with a muon areal density $\rho_\mu > 5.9 \pm 0.4$~m$^{-2}$ and correspond to a rate of  $1.9 \times 10^{-6}$~Hz at the underground location of ALICE. Based upon the number of observed HMM events, the estimated relative statistical uncertainty is 45\%, giving an error in the rate of $\pm 0.9 \times 10^{-6}$~Hz.  

The rate of HMM events obtained with the Monte Carlo can be compared with the observed rate.  
Since we have simulated samples of HMM events corresponding to one year live time, the 
statistical uncertainty in the simulated rate will be lower than that in the measured rate.
Results obtained for the number of HMM events expected in one year from both the simplified Monte Carlo 
and the full simulation (the first of five statistically independent simulations) are shown in the 
first row of Table~\ref{table:hmm-year}.
Comparison of the results demonstrates that the detailed modelling of the 
underground environment has about a 30\% effect on the number of HMM events.  Due to the small 
numbers of HMM events we reused the same simulated EAS sample to perform four additional simulations 
by randomly assigning the core of each shower over the usual surface level area of $205 \times 205$~m$^{2}$.  
Given that the acceptance of the TPC is almost 3000 times smaller, this ensures that the samples are
statistically independent.  A summary of the results obtained for all five simulations is presented
in Table \ref{table:hmm-year} for both CORSIKA 6990 with QGSJET II-03 and CORSIKA 7350
with QGSJET II-04. 

\begin{table}
  \centering
  \begin{tabular}{c|cccc|cccc p{3.0cm}}
  \hline
  {}      &  \multicolumn{4}{c|}{CORSIKA 6990} & \multicolumn{4}{c}{CORSIKA 7350} \\
  {}      &  \multicolumn{4}{c|}{QGSJET II-03} & \multicolumn{4}{c}{QGSJET II-04} \\
  {}      &  \multicolumn{2}{c}{Simple MC} & \multicolumn{2}{c|}{Full MC} & \multicolumn{2}{c}{Simple MC} & \multicolumn{2}{c}{Full MC} \\
  Run & proton          & iron             & proton   & iron        & proton          & iron             & proton   & iron \\ 
  \hline
  \hline
  1 & 40  & 61 & 27 & 51 & 41 & 72 & 30 & 52   \\ 
  2 & 40  & 64 & 24 & 42 & 42 & 88 & 32 & 71   \\
  3 & 31  & 43 & 25 & 31 & 48 & 78 & 29 & 62  \\
  4 & 26  & 52 & 20 & 34 & 46 & 84 & 35 & 61  \\
  5 & 33  & 64 & 22 & 53 & 36 & 83 & 31 & 58   \\
  \hline                               
  \end{tabular}
  \vspace{2ex}
   \caption{Number of HMM events for each run obtained with the simplified Monte Carlo and 
  the full simulation.  Each run is equivalent to 365 days of data taking.
The events have been generated using CORSIKA 6990 with QGSJET II-03 and 
CORSIKA 7350 with QGSJET II-04.}
  \vspace{2ex} 
\label{table:hmm-year}
\end{table}

Final values for the HMM event rate for proton and iron primaries were calculated by taking the average 
value obtained from the five simulations, while the statistical uncertainty was estimated from the standard 
deviation of the 5 values from the mean.  Table \ref{table:hmme_cor6990_cor7350} summarises the mean 
number of HMM events expected in one year for each primary ion calculated from the full simulation. 

\begin{table}
  \centering
  \begin{tabular}{l|cc|cc p{3.0cm}}
  \hline
   {}            & \multicolumn{2}{c|}{CORSIKA 6990} & \multicolumn{2}{c}{CORSIKA 7350}   \\
    {}           & \multicolumn{2}{c|}{QGSJET II-03} & \multicolumn{2}{c}{QGSJET II-04}   \\
               
               & proton         & iron         & proton         & iron    \\ 
   \hline
  \hline 
  $<N>$        & 23.6           & 42.2         & 31.4          & 60.8          \\
  $~~~\sigma$     & 1.3 ($5.5\%$)  & 5.0 ($12\%$) & 1.1 ($3.7\%$) & 3.5 ($5.7\%$) \\
  \hline
  \end{tabular}
  \vspace{2ex}
  \caption{Mean value and statistical uncertainty in the number of HMM events
  for 365 days live time calculated using the full simulation.}
  \label{table:hmme_cor6990_cor7350}          
\end{table}
 
There are two major contributions to the systematic uncertainty on the number of HMM events. 
The first contribution stems from the muon reconstruction algorithm. 
To estimate its contribution we took the first simulated sample, corresponding to 365 days of data taking, 
for each element and each CORSIKA code version and redetermined the number of HMM events using different
tunes of the track selection and matching algorithms.  
The second contribution stems from the uncertainties of the parameters used in the simulations, as discussed 
in section \ref{sec:mmd}.  This was estimated to give an uncertainty in the predicted rate of HMM events 
of approximately $20 \%$. 
Due to the large sample used in the simulations (365 days), the systematic uncertainty is dominant, 
while in the data (30.8 days) the statistical uncertainty is dominant. The systematic uncertainties have been 
added in quadrature to the statistical uncertainty in the final comparison of the observed rate of HMM events 
with that obtained from the Monte Carlo simulations.

\section{Results}

\begin{table}
  \centering
  \begin{tabular}[b]{l|cc|cc|c p{3.0cm}}
  \hline
  {} & \multicolumn{2}{c|}{CORSIKA 6990} & \multicolumn{2}{c|}{CORSIKA 7350} & {}\\
  HMM events & \multicolumn{2}{c|}{QGSJET II-03} & \multicolumn{2}{c|}{QGSJET II-04} & Data \\
  {} & proton & iron & proton & iron & {} \\
  \hline
  \hline
  Period [days per event] & 15.5 & 8.6 & 11.6 & 6.0 & 6.2 \\
  Rate [$\times 10^{-6}$ Hz] & 0.8 & 1.3 & 1.0 & 1.9 & 1.9 \\
  Uncertainty ($\%$) (syst + stat) & 25 & 25 & 22 & 28 & 49 \\
  \hline
  \end{tabular}
  \vspace{2ex}
  \caption{Comparison of the HMM event rate obtained with the full simulation and from measurement.} 
\label{table:ratehmme}
\end{table}

In Table \ref{table:ratehmme} we present the results of this analysis where we compare the rate of simulated HMM events with the measured rate. 
We note that the pure iron sample simulated with CORSIKA 7350 and QGSJET II-04 produces a HMM event rate in close agreement with the measured value. 
The equivalent rate obtained with CORSIKA 6990 and QGSJET II-03 is lower, although still consistent with the measured rate.  
The difference between the two simulations comes primarily from the hadronic model used to generate the EAS.  
It is more difficult to reconcile the measured rate of HMM events with the simulated rate obtained using proton primaries, independent of the version of the model.
However, the large uncertainty in the measured rate prevents us from drawing a firm conclusion about
the origin of these events, although heavy nuclei appear to be the most likely candidates.  
Therefore, an explanation of HMM events in terms of a heavy primary 
cosmic ray composition at high energy and EAS described 
by conventional hadronic mechanisms appears to be 
compatible with our observations. 
This is consistent with the fact that they stem from primaries with 
energies $E > 10^{16}$~eV, where recent measurements  
\cite{kascade_grande_2011, tunka_2015} suggest that the composition 
of the primary cosmic ray spectrum is dominated by heavier elements.

\begin{figure}[!ht]
\centering 
\includegraphics[width=0.8 \textwidth]{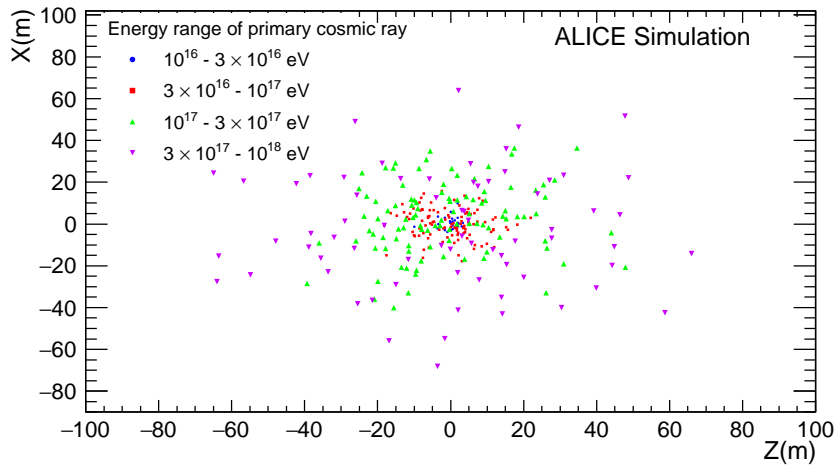}
\caption{The surface level spatial distribution of the cores of simulated EAS giving rise to more than 100 muons in the ALICE Time Projection Chamber.  The simulation was for iron primaries in the energy range $10^{16}-10^{18}$~eV and corresponds to the equivalent of 5 years of data taking. 
}
\label{fig:core_hmme_fe_7350}
\end{figure}

Finally, we have investigated the distribution of simulated EAS core positions at the location of ALICE 
for each of the HMM events simulated with iron primaries using 
CORSIKA 7350 and QGSJET II-04 in Table~\ref{table:hmm-year}, equivalent to 5 years of data taking. 
The distribution is shown in Fig.~\ref{fig:core_hmme_fe_7350}, where the colour of each point indicates the energy 
associated with the primary cosmic ray so as to give a visual representation of the correlation between the 
distance of the core from the centre of ALICE at surface level and the energy of the primary cosmic ray.  We note that the
shower cores of all HMM events fall within an area of approximately $140 \times 140$\ m$^{2}$ 
centred upon ALICE, which is located at the origin in Fig.~\ref{fig:core_hmme_fe_7350}.  
The average distance of the shower core from the centre of ALICE for all events is 19~m and the RMS 
value of the distribution is 16~m. Primaries with an energy $E > 3 \times 10^{17}$~eV, 
corresponding to the highest energy interval studied in this analysis, produce larger showers that may 
give rise to HMM events when the shower core falls farther from the location of ALICE.  
In this case, the mean of the shower core distribution from
the centre of ALICE is 37~m and the RMS value of the distribution is 18~m.

\section{Summary}

In the period 2010 to 2013, ALICE acquired 30.8 days of dedicated cosmic ray data
recording approximately 22.6~million events containing at least one reconstructed muon. 
Comparison of the measured muon multiplicity 
distribution with an equivalent sample of Monte Carlo events suggests a mixed-ion primary 
cosmic ray composition with an average mass that increases with energy. 
This observation is in agreement with most experiments working in the energy 
range of the knee. Following the successful description of the magnitude of the MMD in the 
low-to-intermediate range of muon multiplicities we used the same simulation framework to study the frequency of HMM events.  

High muon multiplicity events were observed in the past by experiments at LEP
but without satisfactory explanation.  Similar high multiplicity events have been observed 
in this study with ALICE. Over the 30.8 days of data taking reported in this paper, 5 events with more 
than 100 muons and zenith angles less than $50^{\circ}$ have been recorded.  
We have found that the observed rate of HMM events is consistent with the rate predicted
by CORSIKA 7350 using QGSJET II-04 to model the development of the resulting air shower,
assuming a pure iron composition for the primary cosmic rays. 
Only primary cosmic rays with an energy $E > 10^{16}$~eV were found to give rise to HMM events.
This observation is compatible with a knee in the cosmic ray energy distribution 
around $3 \times 10^{15}$~eV due to the light component 
followed by a spectral steepening, the onset of which
depends on the atomic number (Z) of the primary.  

The expected rate of HMM events is sensitive to assumptions made about the dominant
hadronic production mechanisms in air shower development.  
The latest version of QGSJET differs from earlier versions in its treatment of forward neutral
meson production resulting in a higher muon yield and has been retuned taking into account
early LHC results on hadron production in 7 TeV proton-proton collisions.
This is the first time that the rate of HMM events, observed at the relatively shallow depth of ALICE, has been satisfactorily reproduced using a conventional hadronic model for the description of extensive air showers; an observation that places significant constraints on alternative, more exotic, production mechanisms.

Compared to the previous studies at LEP, there are two distinguishing aspects of this work that have led to 
these new insights into the origin of HMM events. The first has been the ability to generate large samples 
of very energetic cosmic rays, allowing for a more reliable estimate of the expected rate of these events.
The second, and more important, aspect has been the recent advances in the hadronic description of EAS.
This is a continually evolving field.  We note that in a preparatory study~\cite{alice_ppr2} 
carried out by ALICE in 2004, using an older version of CORSIKA (version 6031), no HMM events
were observed in the MMD distribution simulated for 30 days of data taking with a pure iron primary cosmic 
ray composition.
In the present work, Table \ref{table:ratehmme} gives a quantitative comparison of the rate of HMM events
predicted by two more recent versions of CORSIKA and QGSJET, illustrating the evolution of the hadronic
description of EAS in recent years.  Only in the latest version of the model there 
has been a significant
increase in the rate of HMM events that better approaches the rate observed in this study.

\newpage 
\newenvironment{acknowledgement}{\relax}{\relax}
\begin{acknowledgement}
\section*{Acknowledgements}
The ALICE Collaboration would like to thank all its engineers and technicians for their invaluable contributions to the construction of the experiment and the CERN accelerator teams for the outstanding performance of the LHC complex.
The ALICE Collaboration gratefully acknowledges the resources and support provided by all Grid centres and the Worldwide LHC Computing Grid (WLCG) collaboration.
The ALICE Collaboration acknowledges the following funding agencies for their support in building and
running the ALICE detector:
State Committee of Science,  World Federation of Scientists (WFS)
and Swiss Fonds Kidagan, Armenia,
Conselho Nacional de Desenvolvimento Cient\'{\i}fico e Tecnol\'{o}gico (CNPq), Financiadora de Estudos e Projetos (FINEP),
Funda\c{c}\~{a}o de Amparo \`{a} Pesquisa do Estado de S\~{a}o Paulo (FAPESP);
National Natural Science Foundation of China (NSFC), the Chinese Ministry of Education (CMOE)
and the Ministry of Science and Technology of China (MSTC);
Ministry of Education and Youth of the Czech Republic;
Danish Natural Science Research Council, the Carlsberg Foundation and the Danish National Research Foundation;
The European Research Council under the European Community's Seventh Framework Programme;
Helsinki Institute of Physics and the Academy of Finland;
French CNRS-IN2P3, the `Region Pays de Loire', `Region Alsace', `Region Auvergne' and CEA, France;
German Bundesministerium fur Bildung, Wissenschaft, Forschung und Technologie (BMBF) and the Helmholtz Association;
General Secretariat for Research and Technology, Ministry of
Development, Greece;
Hungarian Orszagos Tudomanyos Kutatasi Alappgrammok (OTKA) and National Office for Research and Technology (NKTH);
Department of Atomic Energy and Department of Science and Technology of the Government of India;
Istituto Nazionale di Fisica Nucleare (INFN) and Centro Fermi -
Museo Storico della Fisica e Centro Studi e Ricerche "Enrico
Fermi", Italy;
MEXT Grant-in-Aid for Specially Promoted Research, Ja\-pan;
Joint Institute for Nuclear Research, Dubna;
National Research Foundation of Korea (NRF);
Consejo Nacional de Cienca y Tecnologia (CONACYT), Direccion General de Asuntos del Personal Academico(DGAPA), M\'{e}xico, :Amerique Latine Formation academique – European Commission(ALFA-EC) and the EPLANET Program
(European Particle Physics Latin American Network)
Stichting voor Fundamenteel Onderzoek der Materie (FOM) and the Nederlandse Organisatie voor Wetenschappelijk Onderzoek (NWO), Netherlands;
Research Council of Norway (NFR);
National Science Centre, Poland;
Ministry of National Education/Institute for Atomic Physics and Consiliul Naţional al Cercetării Ştiinţifice - Executive Agency for Higher Education Research Development and Innovation Funding (CNCS-UEFISCDI) - Romania;
Ministry of Education and Science of Russian Federation, Russian
Academy of Sciences, Russian Federal Agency of Atomic Energy,
Russian Federal Agency for Science and Innovations and The Russian
Foundation for Basic Research;
Ministry of Education of Slovakia;
Department of Science and Technology, South Africa;
Centro de Investigaciones Energeticas, Medioambientales y Tecnologicas (CIEMAT), E-Infrastructure shared between Europe and Latin America (EELA), Ministerio de Econom\'{i}a y Competitividad (MINECO) of Spain, Xunta de Galicia (Conseller\'{\i}a de Educaci\'{o}n),
Centro de Aplicaciones Tecnológicas y Desarrollo Nuclear (CEA\-DEN), Cubaenerg\'{\i}a, Cuba, and IAEA (International Atomic Energy Agency);
Swedish Research Council (VR) and Knut $\&$ Alice Wallenberg
Foundation (KAW);
Ukraine Ministry of Education and Science;
United Kingdom Science and Technology Facilities Council (STFC);
The United States Department of Energy, the United States National
Science Foundation, the State of Texas, and the State of Ohio;
Ministry of Science, Education and Sports of Croatia and  Unity through Knowledge Fund, Croatia.
Council of Scientific and Industrial Research (CSIR), New Delhi, India
\end{acknowledgement}
%
%
%
\bibliography{paper_cosmic_references}{}
\bibliographystyle{utphys}

\newpage

\appendix
%
%
\section{ALICE Collaboration}
\label{app:collab}



\begingroup
\small
\begin{flushleft}
J.~Adam\Irefn{org40}\And
D.~Adamov\'{a}\Irefn{org83}\And
M.M.~Aggarwal\Irefn{org87}\And
G.~Aglieri Rinella\Irefn{org36}\And
M.~Agnello\Irefn{org110}\And
N.~Agrawal\Irefn{org48}\And
Z.~Ahammed\Irefn{org132}\And
S.U.~Ahn\Irefn{org68}\And
S.~Aiola\Irefn{org136}\And
A.~Akindinov\Irefn{org58}
% ...
\renewcommand\labelenumi{\textsuperscript{\theenumi}~}

\section*{Affiliation notes}
\renewcommand\theenumi{\roman{enumi}}
\begin{Authlist}
\item \Adef{0}Deceased
\item \Adef{idp3805088}{Also at: M.V. Lomonosov Moscow State University, D.V. Skobeltsyn Institute of Nuclear, Physics, Moscow, Russia}
\item \Adef{idp5973536}{Also at: University of Kansas, Lawrence, Kansas, United States}
\end{Authlist}

\section*{Collaboration Institutes}
\renewcommand\theenumi{\arabic{enumi}~}
\begin{Authlist}

\item \Idef{org1}A.I. Alikhanyan National Science Laboratory (Yerevan Physics Institute) Foundation, Yerevan, Armenia
\item \Idef{org2}Benem\'{e}rita Universidad Aut\'{o}noma de Puebla, Puebla, Mexico
\item \Idef{org3}Bogolyubov Institute for Theoretical Physics, Kiev, Ukraine
\item \Idef{org4}Bose Institute, Department of Physics and Centre for Astroparticle Physics and Space Science (CAPSS), Kolkata, India
\item \Idef{org5}Budker Institute for Nuclear Physics, Novosibirsk, Russia
\item \Idef{org6}California Polytechnic State University, San Luis Obispo, California, United States
\item \Idef{org7}Central China Normal University, Wuhan, China
\item \Idef{org8}Centre de Calcul de l'IN2P3, Villeurbanne, France
\item \Idef{org9}Centro de Aplicaciones Tecnol\'{o}gicas y Desarrollo Nuclear (CEADEN), Havana, Cuba
\item \Idef{org10}Centro de Investigaciones Energ\'{e}ticas Medioambientales y Tecnol\'{o}gicas (CIEMAT), Madrid, Spain
\item \Idef{org11}Centro de Investigaci\'{o}n y de Estudios Avanzados (CINVESTAV), Mexico City and M\'{e}rida, Mexico
\item \Idef{org12}Centro Fermi - Museo Storico della Fisica e Centro Studi e Ricerche ``Enrico Fermi'', Rome, Italy
\item \Idef{org13}Chicago State University, Chicago, Illinois, USA
\item \Idef{org14}China Institute of Atomic Energy, Beijing, China
\item \Idef{org15}Commissariat \`{a} l'Energie Atomique, IRFU, Saclay, France
\item \Idef{org16}COMSATS Institute of Information Technology (CIIT), Islamabad, Pakistan
\item \Idef{org17}Departamento de F\'{\i}sica de Part\'{\i}culas and IGFAE, Universidad de Santiago de Compostela, Santiago de Compostela, Spain
\item \Idef{org18}Department of Physics and Technology, University of Bergen, Bergen, Norway
\item \Idef{org19}Department of Physics, Aligarh Muslim University, Aligarh, India
\item \Idef{org20}Department of Physics, Ohio State University, Columbus, Ohio, United States
\item \Idef{org21}Department of Physics, Sejong University, Seoul, South Korea
\item \Idef{org22}Department of Physics, University of Oslo, Oslo, Norway
\item \Idef{org23}Dipartimento di Elettrotecnica ed Elettronica del Politecnico, Bari, Italy
\item \Idef{org24}Dipartimento di Fisica dell'Universit\`{a} 'La Sapienza' and Sezione INFN Rome, Italy
\item \Idef{org25}Dipartimento di Fisica dell'Universit\`{a} and Sezione INFN, Cagliari, Italy
\item \Idef{org26}Dipartimento di Fisica dell'Universit\`{a} and Sezione INFN, Trieste, Italy
\item \Idef{org27}Dipartimento di Fisica dell'Universit\`{a} and Sezione INFN, Turin, Italy
\item \Idef{org28}Dipartimento di Fisica e Astronomia dell'Universit\`{a} and Sezione INFN, Bologna, Italy
\item \Idef{org29}Dipartimento di Fisica e Astronomia dell'Universit\`{a} and Sezione INFN, Catania, Italy
\item \Idef{org30}Dipartimento di Fisica e Astronomia dell'Universit\`{a} and Sezione INFN, Padova, Italy
\item \Idef{org31}Dipartimento di Fisica `E.R.~Caianiello' dell'Universit\`{a} and Gruppo Collegato INFN, Salerno, Italy
\item \Idef{org32}Dipartimento di Scienze e Innovazione Tecnologica dell'Universit\`{a} del  Piemonte Orientale and Gruppo Collegato INFN, Alessandria, Italy
\item \Idef{org33}Dipartimento Interateneo di Fisica `M.~Merlin' and Sezione INFN, Bari, Italy
\item \Idef{org34}Division of Experimental High Energy Physics, University of Lund, Lund, Sweden
\item \Idef{org35}Eberhard Karls Universit\"{a}t T\"{u}bingen, T\"{u}bingen, Germany
\item \Idef{org36}European Organization for Nuclear Research (CERN), Geneva, Switzerland
\item \Idef{org37}Excellence Cluster Universe, Technische Universit\"{a}t M\"{u}nchen, Munich, Germany
\item \Idef{org38}Faculty of Engineering, Bergen University College, Bergen, Norway
\item \Idef{org39}Faculty of Mathematics, Physics and Informatics, Comenius University, Bratislava, Slovakia
\item \Idef{org40}Faculty of Nuclear Sciences and Physical Engineering, Czech Technical University in Prague, Prague, Czech Republic
\item \Idef{org41}Faculty of Science, P.J.~\v{S}af\'{a}rik University, Ko\v{s}ice, Slovakia
\item \Idef{org42}Faculty of Technology, Buskerud and Vestfold University College, Vestfold, Norway
\item \Idef{org43}Frankfurt Institute for Advanced Studies, Johann Wolfgang Goethe-Universit\"{a}t Frankfurt, Frankfurt, Germany
\item \Idef{org44}Gangneung-Wonju National University, Gangneung, South Korea
\item \Idef{org45}Gauhati University, Department of Physics, Guwahati, India
\item \Idef{org46}Helsinki Institute of Physics (HIP), Helsinki, Finland
\item \Idef{org47}Hiroshima University, Hiroshima, Japan
\item \Idef{org48}Indian Institute of Technology Bombay (IIT), Mumbai, India
\item \Idef{org49}Indian Institute of Technology Indore, Indore (IITI), India
\item \Idef{org50}Inha University, Incheon, South Korea
\item \Idef{org51}Institut de Physique Nucl\'eaire d'Orsay (IPNO), Universit\'e Paris-Sud, CNRS-IN2P3, Orsay, France
\item \Idef{org52}Institut f\"{u}r Informatik, Johann Wolfgang Goethe-Universit\"{a}t Frankfurt, Frankfurt, Germany
\item \Idef{org53}Institut f\"{u}r Kernphysik, Johann Wolfgang Goethe-Universit\"{a}t Frankfurt, Frankfurt, Germany
\item \Idef{org54}Institut f\"{u}r Kernphysik, Westf\"{a}lische Wilhelms-Universit\"{a}t M\"{u}nster, M\"{u}nster, Germany
\item \Idef{org55}Institut Pluridisciplinaire Hubert Curien (IPHC), Universit\'{e} de Strasbourg, CNRS-IN2P3, Strasbourg, France
\item \Idef{org56}Institute for Nuclear Research, Academy of Sciences, Moscow, Russia
\item \Idef{org57}Institute for Subatomic Physics of Utrecht University, Utrecht, Netherlands
\item \Idef{org58}Institute for Theoretical and Experimental Physics, Moscow, Russia
\item \Idef{org59}Institute of Experimental Physics, Slovak Academy of Sciences, Ko\v{s}ice, Slovakia
\item \Idef{org60}Institute of Physics, Academy of Sciences of the Czech Republic, Prague, Czech Republic
\item \Idef{org61}Institute of Physics, Bhubaneswar, India
\item \Idef{org62}Institute of Space Science (ISS), Bucharest, Romania
\item \Idef{org63}Instituto de Ciencias Nucleares, Universidad Nacional Aut\'{o}noma de M\'{e}xico, Mexico City, Mexico
\item \Idef{org64}Instituto de F\'{\i}sica, Universidad Nacional Aut\'{o}noma de M\'{e}xico, Mexico City, Mexico
\item \Idef{org65}iThemba LABS, National Research Foundation, Somerset West, South Africa
\item \Idef{org66}Joint Institute for Nuclear Research (JINR), Dubna, Russia
\item \Idef{org67}Konkuk University, Seoul, South Korea
\item \Idef{org68}Korea Institute of Science and Technology Information, Daejeon, South Korea
\item \Idef{org69}KTO Karatay University, Konya, Turkey
\item \Idef{org70}Laboratoire de Physique Corpusculaire (LPC), Clermont Universit\'{e}, Universit\'{e} Blaise Pascal, CNRS--IN2P3, Clermont-Ferrand, France
\item \Idef{org71}Laboratoire de Physique Subatomique et de Cosmologie, Universit\'{e} Grenoble-Alpes, CNRS-IN2P3, Grenoble, France
\item \Idef{org72}Laboratori Nazionali di Frascati, INFN, Frascati, Italy
\item \Idef{org73}Laboratori Nazionali di Legnaro, INFN, Legnaro, Italy
\item \Idef{org74}Lawrence Berkeley National Laboratory, Berkeley, California, United States
\item \Idef{org75}Lawrence Livermore National Laboratory, Livermore, California, United States
\item \Idef{org76}Moscow Engineering Physics Institute, Moscow, Russia
\item \Idef{org77}National Centre for Nuclear Studies, Warsaw, Poland
\item \Idef{org78}National Institute for Physics and Nuclear Engineering, Bucharest, Romania
\item \Idef{org79}National Institute of Science Education and Research, Bhubaneswar, India
\item \Idef{org80}Niels Bohr Institute, University of Copenhagen, Copenhagen, Denmark
\item \Idef{org81}Nikhef, Nationaal instituut voor subatomaire fysica, Amsterdam, Netherlands
\item \Idef{org82}Nuclear Physics Group, STFC Daresbury Laboratory, Daresbury, United Kingdom
\item \Idef{org83}Nuclear Physics Institute, Academy of Sciences of the Czech Republic, \v{R}e\v{z} u Prahy, Czech Republic
\item \Idef{org84}Oak Ridge National Laboratory, Oak Ridge, Tennessee, United States
\item \Idef{org85}Petersburg Nuclear Physics Institute, Gatchina, Russia
\item \Idef{org86}Physics Department, Creighton University, Omaha, Nebraska, United States
\item \Idef{org87}Physics Department, Panjab University, Chandigarh, India
\item \Idef{org88}Physics Department, University of Athens, Athens, Greece
\item \Idef{org89}Physics Department, University of Cape Town, Cape Town, South Africa
\item \Idef{org90}Physics Department, University of Jammu, Jammu, India
\item \Idef{org91}Physics Department, University of Rajasthan, Jaipur, India
\item \Idef{org92}Physik Department, Technische Universit\"{a}t M\"{u}nchen, Munich, Germany
\item \Idef{org93}Physikalisches Institut, Ruprecht-Karls-Universit\"{a}t Heidelberg, Heidelberg, Germany
\item \Idef{org94}Purdue University, West Lafayette, Indiana, United States
\item \Idef{org95}Pusan National University, Pusan, South Korea
\item \Idef{org96}Research Division and ExtreMe Matter Institute EMMI, GSI Helmholtzzentrum f\"ur Schwerionenforschung, Darmstadt, Germany
\item \Idef{org97}Rudjer Bo\v{s}kovi\'{c} Institute, Zagreb, Croatia
\item \Idef{org98}Russian Federal Nuclear Center (VNIIEF), Sarov, Russia
\item \Idef{org99}Russian Research Centre Kurchatov Institute, Moscow, Russia
\item \Idef{org100}Saha Institute of Nuclear Physics, Kolkata, India
\item \Idef{org101}School of Physics and Astronomy, University of Birmingham, Birmingham, United Kingdom
\item \Idef{org102}Secci\'{o}n F\'{\i}sica, Departamento de Ciencias, Pontificia Universidad Cat\'{o}lica del Per\'{u}, Lima, Peru
\item \Idef{org103}Sezione INFN, Bari, Italy
\item \Idef{org104}Sezione INFN, Bologna, Italy
\item \Idef{org105}Sezione INFN, Cagliari, Italy
\item \Idef{org106}Sezione INFN, Catania, Italy
\item \Idef{org107}Sezione INFN, Padova, Italy
\item \Idef{org108}Sezione INFN, Rome, Italy
\item \Idef{org109}Sezione INFN, Trieste, Italy
\item \Idef{org110}Sezione INFN, Turin, Italy
\item \Idef{org111}SSC IHEP of NRC Kurchatov institute, Protvino, Russia
\item \Idef{org112}Stefan Meyer Institut f\"{u}r Subatomare Physik (SMI), Vienna, Austria
\item \Idef{org113}SUBATECH, Ecole des Mines de Nantes, Universit\'{e} de Nantes, CNRS-IN2P3, Nantes, France
\item \Idef{org114}Suranaree University of Technology, Nakhon Ratchasima, Thailand
\item \Idef{org115}Technical University of Ko\v{s}ice, Ko\v{s}ice, Slovakia
\item \Idef{org116}Technical University of Split FESB, Split, Croatia
\item \Idef{org117}The Henryk Niewodniczanski Institute of Nuclear Physics, Polish Academy of Sciences, Cracow, Poland
\item \Idef{org118}The University of Texas at Austin, Physics Department, Austin, Texas, USA
\item \Idef{org119}Universidad Aut\'{o}noma de Sinaloa, Culiac\'{a}n, Mexico
\item \Idef{org120}Universidade de S\~{a}o Paulo (USP), S\~{a}o Paulo, Brazil
\item \Idef{org121}Universidade Estadual de Campinas (UNICAMP), Campinas, Brazil
\item \Idef{org122}University of Houston, Houston, Texas, United States
\item \Idef{org123}University of Jyv\"{a}skyl\"{a}, Jyv\"{a}skyl\"{a}, Finland
\item \Idef{org124}University of Liverpool, Liverpool, United Kingdom
\item \Idef{org125}University of Tennessee, Knoxville, Tennessee, United States
\item \Idef{org126}University of the Witwatersrand, Johannesburg, South Africa
\item \Idef{org127}University of Tokyo, Tokyo, Japan
\item \Idef{org128}University of Tsukuba, Tsukuba, Japan
\item \Idef{org129}University of Zagreb, Zagreb, Croatia
\item \Idef{org130}Universit\'{e} de Lyon, Universit\'{e} Lyon 1, CNRS/IN2P3, IPN-Lyon, Villeurbanne, France
\item \Idef{org131}V.~Fock Institute for Physics, St. Petersburg State University, St. Petersburg, Russia
\item \Idef{org132}Variable Energy Cyclotron Centre, Kolkata, India
\item \Idef{org133}Warsaw University of Technology, Warsaw, Poland
\item \Idef{org134}Wayne State University, Detroit, Michigan, United States
\item \Idef{org135}Wigner Research Centre for Physics, Hungarian Academy of Sciences, Budapest, Hungary
\item \Idef{org136}Yale University, New Haven, Connecticut, United States
\item \Idef{org137}Yonsei University, Seoul, South Korea
\item \Idef{org138}Zentrum f\"{u}r Technologietransfer und Telekommunikation (ZTT), Fachhochschule Worms, Worms, Germany
\end{Authlist}
\endgroup

%
\end{document}